\documentclass[11pt,onecolumn,showpacs,showkeys]{revtex4}
\usepackage{epsfig}
\usepackage{graphicx}

\newcommand{\bastar}{\begin{eqnarray*}}
\newcommand{\eastar}{\end{eqnarray*}}
\newskip\humongous \humongous=0pt plus 1000pt minus 1000pt

\newif\ifdtup

\relax
\newcommand{\be}{\begin{equation}}
\newcommand{\ee}{\end{equation}}
\newcommand{\bea}{\begin{eqnarray}}
\newcommand{\eea}{\end{eqnarray}}
\newcommand{\nn}{\nonumber}
\newcommand{\pro}{\partial}

\newcommand{\dfrac}{\displaystyle\frac}
\begin{document}
\titlepage
\title{Dilaton as a Dark Matter Candidate and Its Detection}
\author{Y. M. Cho}
\email{ymcho@yongmin.snu.ac.kr}
\author{J. H. Kim}
\email{rtiger@phya.snu.ac.kr}
\affiliation{Center for Theoretical Physics and School of Physics\\
College of Natural Sciences\\ Seoul National
University, Seoul   151-742, Korea}

\vspace{10mm}

\begin{abstract}
Assuming that the dilaton is the dark matter of the universe,
we propose an experiment to detect the relic dilaton using
the electromagnetic resonant cavity, based on
the dilaton-photon conversion in strong electromagnetic background.
We calculate the density of the relic dilaton, and estimate
the dilaton mass for which the dilaton becomes
the dark matter of the universe. With this we calculate the dilaton
detection power in the resonant cavity, and compare it
with the axion detection power in similar resonant cavity
experiment.
\end{abstract}

\maketitle

\section{Introduction}

One of the important issues in cosmology is the search for the
dark matter. The notable candidates among many dark
matter candidates are the dilaton and axion \cite{prd90,sik}. The two
particles differ completely in their origins, but are very similar
in their coupling to the electromagnetic field and the fermionic
matter fields. The dilaton is a universal scalar field which
appears in all higher-dimensional unified theories
(including the Kaluza-Klein theory and the superstring theory)
which plays the role of the scalar graviton,
and thus couples directly to all matter fields
\cite{jmp75,duff,witt}. On the other hand, the axion is a pseudoscalar
Goldstone boson generated by spontaneous breakdown of the
Peccei-Quinn $U_{PQ}(1)$ symmetry which was introduced to solve
the so called ``strong CP problem" in strong interaction
\cite{pec,wein}. But they have almost identical electromagnetic
coupling, except that the dilation (being a scalar) couples to
$F_{\mu\nu}^2$ while the axion (being a pseudoscalar) couples to
$F_{\mu\nu}{\tilde F}_{\mu\nu}$. In this sense the dilaton and
axion may be viewed as the scalar-pseudoscalar partners of each
other. This is particularly true for the gravitational
axion, the pseudoscalar graviton which has been proposed
by Ni independent of the strong CP problem \cite{ni}.

The axion has been believed to be one of the strong candidates of
the dark matter by many physicists, and experiments to detect
it have been actively performed \cite{sik,asz}. In comparison,
the detection of the dilaton has not so actively been performed
up to now, in spite of its theoretical importance.
It is well known that the dilaton generates the fifth force
which can affect the Einstein's gravity in a fundamental
way \cite{prd87,grg91}. Moreover, in cosmology it can play
the role of the inflaton, and can be an excellent candidate
of the dark matter \cite{prd90,cqg98}.
In this paper we study the dilaton as a candidate
of dark matter in detail, and propose a dilaton detection
experiment using an electromagnetic resonant cavity.
In particular, we refine the existing estimate of
the dilaton mass, calculate the dilaton detection power
in the resonant cavity, and compare this with the axion
detection power in similar experiments.

The paper is organized as follows. In Section II we briefly review
the dilaton physics based on Kaluza-Klein theory.
In particular, we discuss how the dilaton mass can resolve
the hierarchy problem and determine the size of the internal space.
In Section III we discuss the role of dilaton in cosmology, and
estimate the number density of the relic dilaton in the present
universe based on the dilaton decay to two photons and
fermion-antifermion pairs. In Section IV we discuss the condition
for the dilaton to be a candidate of dark matter, and refine the
acceptable mass range of dilaton. In Section V we propose the
experiment to detect the dilaton using an electromagnetic
resonant cavity. We calculate the dilaton detection
power in the resonant cavity, and compare it with
the axion detection power in similar experiments. Finally in
Section VI we discuss the physical implications of our analysis.

\section{Dilatonic Fifth Force and Hierarchy Problem}

All known interactions are mediated by spin-one or spin-two fields.
However, the unification of all interactions inevitably requires
the existence of a fundamental spin-zero field. In fact, all modern
unified theories (Kaluza-Klein theory, supergravity, and
superstring) contain a fundamental scalar field called the dilaton,
or more precisely the Kaluza-Klein dilaton \cite{jmp75,prd87}.
What makes this scalar field unique is that unlike
others scalar fields like the Higgs field, it couples directly
to the (trace of the) energy-momentum tensor of the matter fields.
As such it plays the role of the scalar graviton, and
generates the dilatonic fifth force which modifies Einstein's
gravity in a fundamental way.

Actually the simplest unified theory which contains the dilaton
is the Brans-Dicke theory \cite{bra,prl92}.
Unfortunately the Brans-Dicke dilaton is proposed as a massless
scalar graviton, so that it must create a long
range fifth force which is comparable to Newton's gravitational
force. This contradicts the experiments
which tell that such long range fifth force does not
exist in nature \cite{exp1,exp2}. This rules out
the Brans-Dicke theory as unphysical.
On the other hand, the Kaluza-Klein dilaton has
no such problem, because it naturally acquires a mass
and generates a short range fifth force which does not
contradict all known experiments \cite{jmp75,prd87}.
So we will discuss the Kaluza-Klein dilaton in detail
in the following.

The Kaluza-Klein dilaton plays a crucial role to resolve
the so called hierarchy problem. It has been very difficult
to understand why the Planck mass fixed by
the Newton's constant is so large compared to the mass scale of
ordinary elementary particles, or equivalently
why the gravitational force is so weak compared to other forces.
There have been many proposals to resolve this problem.
Long time ago Dirac conjectured that the Newton's constant
may not be a constant but actually a time-dependent
parameter to resolve the problem \cite{dirac}.
Another proposal based on the higher-dimensional
unification is that the gravitational force
in higher-dimension is actually as strong as
other forces, but a relatively large (compared to the Planck scale)
internal space of the order of TeV scale makes
the 4-dimensional gravitational force very weak \cite{ark,dim}.
In this section we show that the dilaton plays
the pivotal role in both proposals to resolve
the hierarchy problem.

Since all higher-dimensional unified theories contain
the $(4+n)$-dimensional gravitation, we start from
the Kaluza-Klein theory. To obtain the $4$-dimensional
effective theory one has to make the dimensional reduction.
A simple and elegant way to do this is to impose
an isometry \cite{jmp75,pen}.
In this dimensional reduction by isometry one may view
the $(4+n)$-dimensional unified space as a principal fiber
bundle P(M,G) made of the $4$-dimensional space-time
manifold M as the base manifold and
$n$-dimensional group manifold G as the vertical fiber
(the internal space) on which G acts as an isometry group.
Let $\gamma_{\mu\nu}$ and $\tilde{\phi}_{ab}$ be
the 4-dimensional metric on M and the n-dimensional
metric on G, $\gamma$ and $\tilde{\phi}$ be the determinants
of $\gamma_{\mu\nu}$ and $\tilde{\phi}_{ab}$, and
$\rho_{ab}=\tilde{\phi}_{ab}/\root n\of{\tilde{\phi}}$
($|\textrm{det}\rho_{ab}|=1$) be the normalized metric on G.
In this setting the (4+n)-dimensional
Einstein-Hilbert action on P leads to the following 4-dimensional
Lagrangian in the Jordan frame \cite{jmp75}
\bea
&\mathcal{L}_{CF}= -\dfrac{\hat V_G}{16\pi G_P}
\sqrt{\vphantom{\tilde{\phi}}\gamma}\sqrt{\tilde\phi}
~\Bigg[~~R_M-\dfrac{n-1}{4n}\gamma^{\mu\nu}\dfrac{(\partial_{\mu}
\tilde{\phi})(\partial_{\nu}\tilde{\phi})}{\tilde{\phi}^{2}}
+\dfrac{\kappa^2}{4} \root n\of{\tilde{\phi}}
~\rho_{ab}\gamma^{\mu\alpha}\gamma^{\nu\beta}
F_{\mu\nu}^{a}F_{\alpha\beta}^b \nn\\
&+\dfrac{\gamma^{\mu\nu}}{4}(D_\mu \rho^{ab})(D_\nu \rho_{ab})
+\dfrac{1}{\kappa^2 \root n\of{\tilde \phi}}~\hat R_G(\rho_{ab})
+\Lambda_P+\lambda(|\textrm{det}\rho_{ab}|-1)~\Bigg],
\label{cflag1}
\eea
where $G_P$ is the (4+n)-dimensional Newton's constant,
$\hat V_G$ is the normalized volume of the internal space G,
$R_M$ is the scalar curvature of M fixed by $\gamma_{\mu\nu}$,
$\hat R_G(\rho_{ab})$ is the normalized
internal curvature fixed by $\rho_{ab}$,
$\kappa$ is the unit scale of the internal space G,
$F_{\mu\nu}^{a}$ is the gauge field of the isometry group G,
$\Lambda_P$ is a (4+n)-dimensional cosmological constant,
and $\lambda$ is a Lagrange multiplier.

Notice that the scalar field $\tilde{\phi}$ couples non-minimally
to $R_M$, so that the metric $\gamma_{\mu\nu}$ does not describe
the massless spin-two graviton \cite{prl92}. To cure this defect
and discuss the physics of (\ref{cflag1}), we have to choose the
physical conformal frame in which the metric describes the
massless spin-two graviton. Let $<\tilde \phi>={v_0}^2$ and
introduce the Pauli metric $g_{\mu\nu}$ and the Kaluza-Klein
dilaton $\sigma$ by
\bea
g_{\mu\nu}= \exp \Big(\sqrt{\dfrac{n}{n+2}}~\sigma \Big)
\gamma_{\mu\nu}~, ~~~~~~~\tilde{\phi}= \Big[v_0~\exp
\Big(\sqrt{\dfrac{n}{n+2}}~\sigma \Big) \Big]^2.
\eea
The reason why we call $\sigma$ the dilaton is obvious. It
determines the local dilatation in the conformal transformation. 
With this we find the following Lagrangian in the Pauli
frame \cite{prd90,prd87},
\bea
&\mathcal{L}_{CF}= -\dfrac{v_0\hat V_G}{16\pi G_P} \sqrt{g}
\Big[R+\frac{1}{2}(\partial_{\mu}\sigma)^2
-\dfrac{1}{4}(D_\mu \rho^{ab})(D^{\mu}{\rho}_{ab}) \nn\\
&+\kappa^{-2} {v_0}^{-2/n} \hat R_G(\rho_{ab}) \exp
\Big(-\sqrt{\dfrac{n+2}{n}}~\sigma \Big)
+\Lambda_P \exp \Big(-\sqrt{\dfrac{n}{n+2}}~\sigma \Big) \nn\\
&+\lambda\exp \Big(-\sqrt{\dfrac{n}{n+2}}\sigma\Big)
(|\textrm{det}\rho_{ab}|-1) +\dfrac{\kappa^2}{4} {v_0}^{2/n} \exp
\Big(\sqrt{\dfrac{n+2}{n}}~\sigma \Big)
\rho_{ab}F_{\mu\nu}^{a}F^{\mu\nu b} ~\Big] \nn\\
&=-\dfrac{\sqrt{g}}{16\pi G}
\Big[R+\frac{1}{2}(\partial_{\mu}\sigma)^2
-\dfrac{1}{4}(D_\mu \rho^{ab})(D^{\mu}{\rho}_{ab}) \nn\\
&+\dfrac{1}{16\pi G} {v_0}^{-2/n} \hat R_G(\rho_{ab}) \exp
\Big(-\sqrt{\dfrac{n+2}{n}}~\sigma \Big)
+\Lambda_P \exp \Big(-\sqrt{\dfrac{n}{n+2}}~\sigma \Big) \nn\\
&+\lambda\exp \Big(-\sqrt{\dfrac{n}{n+2}}\sigma\Big)
(|\textrm{det}\rho_{ab}|-1)
+4\pi G \exp \Big(\sqrt{\dfrac{n+2}{n}}~\sigma \Big)
\rho_{ab} {\hat F}_{\mu\nu}^{a} {\hat F}^{\mu\nu b}~\Big],
\label{cflag2}
\eea
where we have put
\bea
&\dfrac{v_0\hat V_G}{16\pi G_P}=\dfrac{1}{16\pi G},
~~~~~\dfrac{\kappa^2}{16\pi G}=1,
\eea
and renormalized the field strength $F_{\mu\nu}^{a}$ to ${\hat
F}_{\mu\nu}^{a}={v_0}^{1/n} F_{\mu\nu}^{a}$ to assure the minimal
coupling of the Pauli metric to the gauge field.

Notice that the unit scale of the internal space $\kappa$ is fixed
by the Planck scale $\sqrt{16\pi G}$, but the actual scale of the
internal space is given by ${v_0}^{1/n} \kappa$, because the
vacuum expectation value of the volume of the internal space is
fixed by
\bea
&<V_G>=\sqrt {<\tilde \phi>}~\hat V_G =v_0~\hat V_G \simeq
v_0~\kappa^n\simeq v_0~(16\pi G)^{n/2}.
\eea
This tells that the scale of the higher-dimensional
gravitational constant $G_P$ need not be fixed by the
Planck scale, because it is given by \cite{prd90,grg91}
\bea
&{G_P}^{1/(n+2)}=(16\pi)^{n/2(n+2)}~{v_0}^{1/(n+2)}~G^{1/2}.
\label{veq}
\eea
So, with a large $v_0$, one can easily bring the length scale
$\root (n+2)\of{G_P}$ to the order of the elementary particle 
length scale. Indeed,
$n=2$ brings the Plank scale down to TeV scale when the scale of
the internal space becomes of the order of $10^{-1} \rm{cm}$. This
is precisely the proposal which has been popularized to resolve
the hierarchy problem \cite{ark,dim}.

Now we show how the dilaton can resolve the hierarchy
problem. Consider the gravitational coupling to
the gauge field in (\ref{cflag2}). Here the dilaton modifies
$G$ to $G\exp[\sqrt{(n+2)/n}~\sigma]$,
which can be interpreted as a space-time dependent
Newton's constant. So the dilaton transforms
the hierarchy problem to
a space-time dependent artifact \cite{jmp75,prd87}.
And this is precisely the Dirac's proposal to resolve
the hierarchy problem. Furthermore, with a large internal space,
we can show that the dilaton can bring down the Planck mass
to the ordinary elementary particle mass.
To see this, suppose the Lagrangian
(\ref{cflag2}) has the unique vacuum at
\bea
&<g_{\mu\nu}>=\eta_{\mu\nu},~~~<\sigma>=0,
~~~<\rho_{ab}>=\delta_{ab},~~~<A_\mu^a>=0.
\eea
Then we have the following dilatonic potential
$V(\sigma)$ \cite{prd87,prd90}
\bea
&V(\sigma)=\dfrac{1}{16\pi G} \Big[\dfrac{\hat R_G}{16\pi G}
~{v_0}^{-2/n}~\exp (-\sqrt{\dfrac{n+2}{n}}~\sigma)
+\Lambda_P~\exp (-\sqrt{\dfrac{n}{n+2}}~\sigma) \Big] +V_0 \nn\\
&=\dfrac{\hat R_G}{(16\pi G)^2}~{v_0}^{-2/n} ~\Big[\exp
(-\sqrt{\dfrac{n+2}{n}}~\sigma \big) -\dfrac{n+2}{n}~\exp
(-\sqrt{\dfrac{n}{n+2}}~\sigma) +\dfrac{2}{n} \Big], \label{dpot}
\eea
where $\hat R_G=\hat R_G(<\delta_{ab}>)$ is the dimensionless
vacuum curvature of the internal space $G$ obtained by the
bi-invariant Cartan-Killing metric $\delta_{ab}$
\bea
&\hat R_G=-\dfrac12 f_{ab}^{~~d}f_{cd}^{~~b} \delta^{ac}
-\dfrac 14 f_{ab}^{~~m}f_{cd}^{~~n}\delta^{ac}\delta^{bd}\delta_{mn},
\label{intc}
\eea
and $V_0$ is a constant which assures that (\ref{dpot})
does not create non-vanishing 4-dimensional cosmological constant
(vacuum energy).
An important point here is that $\Lambda_P$ and $V_0$
are completely fixed by the vacuum condition
$dV(0)/d\sigma=0$ and $V(0)=0$,
\bea
&\Lambda_P=-\dfrac{n+2}{n}~{v_0}^{-2/n} ~\dfrac{\hat R_G}{16\pi
G}, ~~~~~V_0=\dfrac{2}{n}~{v_0}^{-2/n} ~\dfrac{\hat R_G}{(16\pi
G)^2}.
\eea
With this we find the following mass
$\mu$ of the Kaluza-Klein dilaton,
\bea
&\mu^2= (16\pi G)~\dfrac{d^2 V(0)}{d\sigma^2}
=-{v_0}^{-2/n}~\dfrac{\hat R_G}{8\pi n~G}
=-{v_0}^{-2/n}~\dfrac{\hat R_G}{8\pi n}~m_p^2, \label{dmass}
\eea
where $m_p\simeq 1.2\times 10^{19}~\mathrm{GeV}$ is the Planck
mass. This confirms that when the internal space is of the Planck
scale (i.e., when $v_0 \simeq 1$) the dilaton mass becomes of the
Planck mass. But remarkably, a large $v_0$ naturally reduces the
dilaton mass to the order of the elementary particle mass scale
when $\hat R_G \neq 0$ \cite{prd90,grg91}. In fact (\ref{dmass})
tells that the dilaton mass is determined by the scale of the
internal space $L_G$ as follows,
\bea
L_G={v_0}^{1/n}~\kappa=\sqrt{-\dfrac{2\hat R_G}n} ~\dfrac{1}{\mu}
\simeq \dfrac{1}{\mu}. \label{lg}
\eea
In particular, for the $S^3$ compactification of the
$3$-dimensional internal space in $(4+3)$-dimensional unification
with G=$SU(2)$, we have $\hat R_G=-3/2$ and $L_G=1/\mu$.
This is how the dilaton resolves the hierarchy problem
in Kaluza-Klein unification.

At this point it is important to compare (\ref{veq}) and
(\ref{dmass}). Both provide a resolution of the hierarchy problem,
but there are important differences. First, (\ref{veq}) does that
with the gravitational coupling strength, while (\ref{dmass}) does
that with the dilaton mass. Secondly, the dimension of the
internal space $n$ plays the crucial role in (\ref{veq}), but the
curvature of the internal space plays the crucial role in
(\ref{dmass}). In fact we have $\mu=0$ when $\hat R_G=0$,
independent of $n$ and $v_0$. More importantly, (\ref{dmass})
tells that a mass can be generated geometrically through the
scalar curvature of the internal space \cite{prd87,prd90}. This
demonstrates that there is another mass generation mechanism other
than the Higgs mechanism, a geometric mass generation through the
curvature of space-time. Understanding the origin of mass has been
a fundamental problem in physics. Our analysis shows that the
hierarchy problem is closely related to the problem of the origin
of mass, and that the geometric mass generation provides a natural
resolution to the problem of the origin of mass.

In superstring or supergravity unification the situation is
similar but more complicated, because in this case one has other
higher-dimensional matter fields \cite{prd87,cqg98}.
For example, in superstring one has an extra
higher-dimensional dilaton (the string dilaton) which remains
massless in all orders of perturbation, so that one has to
find out a natural mechanism to make the dilaton massive
first \cite{witt}. Other than these complications
the generic features of the dilaton physics remain the same.
This makes the dilaton a fundamental scalar field of nature
which one can not ignore.

The dilaton has been called in various names, recently by the
radion \cite{ark} or the chameleon \cite{kho}. But we notice
that the dilaton as the scalar graviton has a long history. The
first such dilaton was the Brans-Dicke dilaton introduced by
Jordan and independently by Brans and Dicke \cite{bra}.
Subsequently the Kaluza-Klein dilaton \cite{jmp75} and string
dilaton \cite{witt} have been introduced. Later, the dilaton 
has been re-invented by many authors in so-called 
``the scalar field models". Among these only the
Kaluza-Klein dilaton naturally acquires the mass and thus can
describe a realistic scalar graviton.

As we have remarked an immediate consequence of the dilaton is the
presence of dilatonic fifth force which modifies Einstein's
gravitation \cite{prl92}. To see how the dilaton affect the
gravitation we have to know the mass of dilation and its coupling
strength to matter fields. In Kaluza-Klein theory
the dilaton naturally acquires a mass as we
have shown in (\ref{dmass}). As for the dilatonic coupling to matter
fields, the coupling may depend on the types of matter field
it couples to \cite{prd87,prl92}.
But in practice only one type of coupling, the
dilatonic coupling to the baryonic matter, is important because
this is what we measure in experiments. So, only two parameters,
the baryonic coupling constant and the mass of the dilaton,
becomes important to describe the dilatonic fifth-force.
Let $F_g$ and $F_5$ be the gravitational and the fifth force
between the two baryonic point particles
separated by a distance $r$. From the dimensional argument, one
may express the total force in the Newtonian limit as
\bea \label{Tforce}
F=F_{g}+F_{5} \simeq\frac{\alpha_{g}}{r^{2}}
+\dfrac{\alpha_{5}}{r^{2}}~e^{-\mu r}
=\dfrac{\alpha_{g}}{r^{2}} \big(1+\beta~e^{-\mu r} \big)
\eea
where $\alpha_g$, $\alpha_5$ are the fine structure constants of
the gravitation and fifth force, and $\beta$ is the ratio between
them. In terms of Feynman diagrams the first term represents one
graviton exchange but the second term represents one dilaton
exchange in the zero momentum transfer limit. In the Kaluza-Klein
unification we have $\beta=n/(n+2)$ \cite{prd87,prd90},
but in general one may assume $\beta \simeq 1$ because
the dilaton is the scalar partner of the graviton.
With this assumption one may try to measure the range of
the fifth force experimentally.

A recent torsion-balance fifth force experiment puts
the upper bound of the range of the fifth force
to be around $56~\mathrm{\mu m}$ with 95\% confidence
level \cite{exp1,exp2}. This tells that the
dilaton mass has to be larger than $10^{-2}~\mathrm{eV}$.
This, with (\ref{lg}), implies that,
in the $(4+3)$-dimensional unification with the $S^3$
compactification of the internal space, the scale of
the internal space $L_G$ is smaller than $44~ \mathrm{\mu m}$. In
the following, however, we will simply treat the dilaton mass an
undetermined parameter, and find an independent estimate of the
dilaton mass based on the assumption that the dilaton is the dark
matter of the universe.

\section{Relic Dilaton in Cosmology}

The dilaton has another important impact in cosmology. First of
all, it could be a natural candidate for the dark matter of the
universe \cite{prd90,cqg98}. The dilaton starts with the thermal
equilibrium at the beginning and decouples from other sources very
early near the Planck time. Moreover, since its coupling to matter
fields is very weak, it may easily survive in the present universe
and become the dark matter of the universe. In this section we
estimate the density of the relic dilaton.

Let's consider the dilaton in the early universe. From
the dimensional argument one may assume the
dilatonic coupling strength to matter fields to be $g~m/m_{p}$,
where $g$ is the dimensionless coupling constant and
$m$ is the mass of the relevant matter (e.g.,
quarks and gluons).
But at high temperature (at $T \gg m$), the coupling strength
can be written as $g~T/m_{p}$. With this one can
easily estimate the dilaton creation (and annihilation) cross
section as \cite{cqg98}
\bea
\sigma \simeq g^2~\Bigg(\frac{T}{m_{p}}\Bigg)^{2}
\times \frac{1}{T^2},
\eea
with the transition rates $\Gamma$
\bea
\Gamma \simeq N \sigma v  \simeq
g^2~\Bigg(\frac{T}{m_{p}}\Bigg)^{2} \times T,
\eea
where $N$ and $v$ are the density of the matter and the speed
of the dilaton. Similarly the dilaton scattering cross section
and the interaction rate are given by
\bea
\sigma \simeq g^4~\Bigg(\frac{T}{m_{p}}\Bigg)^{4}
\times \frac{1}{T^2},
~~~~~~\Gamma \simeq N \sigma v  \simeq
g^4~\Bigg(\frac{T}{m_{p}}\Bigg)^{4} \times T.
\eea
On the other hand, the Hubble expansion rate in the early
universe is given by $H \simeq T^{2}/m_{p}$.
So, letting $ \Gamma \simeq H $ we find the dilaton
decoupling temperature
\begin{equation}
T_{D} \simeq \dfrac{m_{p}} {g^{4/3}}.
\end{equation}
This confirms that the dilaton is thermally produced at
the beginning, and decouples from the other matters around
the Planck time.

The dilaton becomes unstable and decays into ordinary matter.
A typical decay process
is the two-photon process and the
fermion-antifermion pair production process.
The Lagrangian (\ref{cflag2}) implies that,
in the linear approximation where $\sigma$
is assumed small enough, the decay may be described
by the following interaction Lagrangian,
\begin{equation}
\mathcal{L}_{int} \simeq -\frac{1}{4} g_{1} \sqrt{16 \pi G}
~\hat{\sigma}~F_{\mu\nu}F^{\mu\nu} - g_{2} \sqrt{16 \pi G}~m
~\hat{\sigma}~\bar{\psi}\psi,
\label{dmint}
\end{equation}
where $g_1$ and $g_2$ are dimensionless coupling constants,
$m$ is the mass of the fermion, and $\hat \sigma=\sigma/\sqrt{16\pi G}$
is the dimensional (physical) dilaton field.
This should be compared to
the following axion interaction Lagrangian given by \cite{sik,ni},
\begin{equation}
\mathcal{L}_{int} \simeq - \alpha_{\gamma}~a
~F_{\mu\nu}\tilde{F}^{\mu\nu}
-i\alpha_{f}~a~\bar{\psi}\gamma^{5}\psi,
\end{equation}
where $a$ is the axion field, $\alpha_{\gamma}$ and $\alpha_f$ are
the axion coupling constants. This confirms that dilaton and axion
are the scalar-pseudoscalar counterparts of each other.
Actually we can also include the following
dilaton-fermion interaction in (\ref{dmint}) 
\bea
&g_3 \sqrt{16 \pi G}~\hat{\sigma}
~\bar{\psi} \gamma^\mu \pro_\mu \psi 
+g_4 \sqrt{16 \pi G}~\pro_\mu \hat{\sigma}
~\bar{\psi} \gamma^\mu \psi.
\eea
But for simplicity we will concentrate on (\ref{dmint}) 
in the following.

Consider the interaction between dilaton and photon first,
and let's introduce a dimensional coupling constant
$\alpha_1=g_{1}\sqrt{16 \pi G}/4$ and denote the
dilaton mass by $\mu$. The differential dilaton decay rate to
two photons at tree level is given by
\bea
&d\Gamma_{\sigma\rightarrow\gamma\gamma}
=\dfrac{1}{2p^{0}}\sum_{\lambda, \lambda'=\pm1}
\dfrac{1}{2!}(2\pi)^{4}\delta^{(4)}(p^{\mu}-k^{\mu}-k'^{\mu})|M|^{2}
\dfrac{d^{3}\vec{k}}{(2\pi)^{3}2k^{0}}
\dfrac{d^{3}\vec{k'}}{(2\pi)^{3}2k'^{0}}~,\\
&M=-i\alpha_1 \Big(k_{\mu}\epsilon_{\nu}(k,\lambda)
-k_{\nu}\epsilon_{\mu}(k,\lambda)\Big)
\Big(k'^{\mu}\epsilon'^{\nu}(k',\lambda')
-k'^{\nu}\epsilon'^{\mu}(k',\lambda')\Big),
\eea
where $p^{\mu}$ and $k^{\mu},~k'^{\mu}$ are the 4-momenta of the incoming
dilaton and the outgoing photons, $M$ is the reduced Feynman
matrix element, $\epsilon^{\mu}(k,\lambda)$ and
$\epsilon'^{\mu}(k',\lambda')$ are the transverse polarization
vectors of photons. It is simple to calculate the matrix element in the
center of momentum (COM) frame where
$\vec{k'}=-\vec{k}$. From
\bea
&\Big(k_{\mu}\epsilon_{\nu}(k,\lambda)-k_{\nu}
\epsilon_{\mu}(k,\lambda)\Big)\Big(k'^{\mu}\epsilon'^{\nu}(k',\lambda')
-k'^{\nu}\epsilon'^{\mu}(k',\lambda')\Big)
=2\Big(k_{\mu}k'^{\mu}\Big)\Big(\epsilon_{\nu}(k,\lambda)
\epsilon'^{\nu}(k',\lambda')\Big),  \nn\\
&k_\mu k'^\mu=-2|\vec{k}|^2,
~~~~~\dfrac{}{}\sum_{\lambda,\lambda'}\Big|\epsilon_{\nu}(k,\lambda)
\epsilon'^{\nu}(k',\lambda')\Big|^2=2,
\eea
we get the following decay rate,
\bea
&\Gamma_{\sigma\rightarrow\gamma\gamma}
= \dfrac{\alpha_1^2}{2\pi^2\mu} \int d^3 \vec{k}~ d^3
\vec{k'} |\vec{k}|^2 \delta^{(4)}(k^\mu+k'^\mu-p^\mu) \nonumber \\
&=\dfrac{\alpha_1^2}{2\pi^2 \mu}\int d^3 d^3 \vec{k'}
\vec{k}~|\vec{k}|^2 \delta(k^0+k'^0-\mu)
\delta^{(3)}(\vec{k}+\vec{k'})
=\dfrac{\alpha_1^2}{2\pi^2 \mu}\int d^3 \vec{k}~|\vec{k}|^2
\delta(2k^0-\mu) \nn\\
&=\dfrac{\alpha_1^2}{2\pi^2 \mu}\int d|\vec{k}|d\Omega_{\vec{k}}
|\vec{k}|^4 \frac{1}{2}\delta(|\vec{k}|-\mu/2)
=\dfrac{\alpha_1^2 \mu^3}{16\pi}.
\eea
With this we get the following life-time of the dilaton
\begin{equation}
\tau_{\sigma\rightarrow\gamma\gamma}
=\dfrac{1}{\Gamma_{\sigma\rightarrow\gamma\gamma}}
=\dfrac{16 m_{p}^2}{g_1^2\mu^3}.
\label{plt}
\end{equation}
Notice that when $\mu\simeq m_p$, the dilaton has a very short life-time.

Now consider the dilaton-fermion interaction,
and let $\alpha_2=g_2 \sqrt{16\pi G}~m$ be the dimensionless
coupling constant.
The differential decay rate of dilaton to
fermion and anti-fermion pair at tree level is written as
\bea
&d\Gamma_{\sigma\rightarrow\bar{\psi}\psi}
=\dfrac{1}{2p^{0}}\sum_{s,s'=\pm \frac{1}{2}}
(2\pi)^{4}\delta^{(4)}(p^{\mu}-k^{\mu}-k'^{\mu})|M|^{2}
\dfrac{d^{3}\vec{k}}{(2\pi)^{3}2k^{0}}
\dfrac{d^{3}\vec{k'}}{(2\pi)^{3}2k'^{0}},  \nn\\
&M=-i\alpha_2\bar{u}(k,s) \textit{v}(k',s'),
\eea
where $p^{\mu}$ and $k^{\mu}$, $k'^{\mu}$ are the 4-momenta of the
incoming dilaton and the outgoing fermion-antifermion pair,
and $s,~s'$ are the fermion spin
indices. Using the well-known sum-rule \cite{gre},
\begin{equation}
\sum_{s,s'=\pm\frac{1}{2}}|\bar{u}(k,s)
\textit{v}(k',s')|^2=4(-k_\mu k'^\mu-m^2),
\end{equation}
we have the following decay rate,
\bea
&\Gamma_{\sigma\rightarrow\bar{\psi}\psi}
=\dfrac{\alpha_2^2}{8\pi^2 p^0}\int \dfrac{d^3 \vec{k}}{k^0}~
\dfrac{d^3 \vec{k'}}{k'^0}(-k_\mu k'^\mu-m^2)
\delta^{(4)}(k^\mu+k'^\mu-p^\mu) \nonumber \\
&=\dfrac{\alpha_2^2}{8\pi^2 p^0}\times\dfrac{\mu^2-4m^2}{2}
\int \dfrac{d^3 \vec{k}}{k^0}~\dfrac{d^3 \vec{k'}}{k'^0}
\delta^{(3)}(\vec{k}+\vec{k'}-\vec{p})\delta(k^0+k'^0-p^0)  \nonumber \\
&= \dfrac{\alpha_2^2}{8\pi^2 \mu}\times\dfrac{\mu^2-4m^2}{2}
\int \dfrac{d^3 \vec{k}}{(k^0)^2} \delta(2k^0-\mu)
~~(\textrm{COM frame})  \nonumber \\
&= \dfrac{\alpha_2^2}{8\pi^2 \mu}\times\dfrac{\mu^2-4m^2}{2}
\int d|\vec{k}|d\Omega_{\vec{k}}\dfrac{|\vec{k}|^2}{(k^0)^2}
\delta(2\sqrt{m^2+|\vec{k}|^2}-\mu)  \nonumber \\
&= \dfrac{\alpha_2^2}{2\pi \mu}\times\dfrac{\mu^2-4m^2}{2}
\times\dfrac{|\vec{k}|}{2(k^0)}\Bigg|_{k^0=\mu/2}
= \dfrac{\alpha_2^2\mu}{8\pi}\times
\Bigg[1-\bigg(\dfrac{2m}{\mu}\bigg)^2\Bigg]^{3/2}.
\eea
So we have the following life-time of the dilaton
\begin{equation}
\tau_{\sigma\rightarrow\bar{\psi}\psi}
=\frac{1}{\Gamma_{\sigma\rightarrow\bar{\psi}\psi}}
=\frac{m_{p}^2}{2g_{2}^2
m^2\mu}\Bigg[1-\bigg(\frac{2m}{\mu}\bigg)^2\Bigg]^{-3/2}.
\label{flt}
\end{equation}
Notice that this becomes comparable to (\ref{plt}) only when
$m\sim 0.32\times\mu$, so that the two photon decay becomes the
dominant decay of dilaton in general.

The dilaton number density $n$ after the decoupling
is given by the well-known equation \cite{kol}.
\bea
&\dfrac{d(nR^3)}{dt}=-\dfrac{1}{\tau}(nR^3),
~~~~~\dfrac{dn}{dt}+3H~n=-\dfrac1{\tau}~n.
\eea
where $\tau$ is the total life-time, $R$ is the scale factor
of the Friedmann-Robertson-Walker metric, and $H$ is the Hubble parameter.
From this we have the familiar expression
\begin{equation}
n(t)=n_D\Bigg(\dfrac{R_{D}}{R}\Bigg)^3 \textrm{exp}(-t/\tau),
\end{equation}
where the subscript $D$ denotes the decoupling time.
Note that the factor $1/R^3$ represents the dilution of the dilaton
due to Hubble expansion. To find the present
dilaton number density notice that in the highly
relativistic regime (i.e., when $T\gg \mu$), the
particle number density is given by \cite{kol},
\bea
&n_b = \dfrac{\zeta(3)}{\pi^2}gT^3 ~~~\textrm{(for a boson)},  \nn\\
&n_f = \dfrac{3}{4} \dfrac{\zeta(3)}{\pi^2}gT^3 ~~~\textrm{(for a
fermion)},
\eea
where $g$ is the internal degrees of freedom of the relevant
particle and $\zeta(x)$ is the Riemann's zeta function.
So, at the time of dilaton decoupling, the dilaton
number density $n_D$ is given by
\begin{eqnarray}
n_D &=& \dfrac{\zeta(3)}{\pi^2}~T_{D}^3
\simeq \dfrac{1.202}{\pi^2}~T_{D}^3.
\end{eqnarray}
On the other hand, the total entropy density $s$ of the universe
is given by \cite{kol},
\begin{eqnarray}
&s = \dfrac{2\pi^2}{45}g_{*}T^3, \nn\\
&g_{*}=\dfrac{}{}\sum_{i=bosons}g_i \Bigg
(\dfrac{T_i}{T}\Bigg)^3+\frac{7}{8}\sum_{i=fermions}g_i
\Bigg(\dfrac{T_i}{T}\Bigg)^3,
\end{eqnarray}
where $g_i$ and $T_i$ are the internal degrees of freedom and
the thermal equilibrium temperature of the $i$-th particle,
and $T$ is the thermal temperature of
photon. At present we have $g_{*0}\simeq 3.91$ (with photon
and three types of light neutrinos), but at the
Plank time we have $g_{*}\simeq 106.75 $ according to the
standard model \cite{kol}. Now, the total
entropy conservation of the universe in the co-moving volume
tells that ${g_*}_{D}T_{D}^3 R_{D}^3 = {g_*}_{0}T_{0}^3 R_{0}^3$.
From this we get (with $T_0\simeq 2.73~\textrm{K}$)
the present dilaton number density $n(t_{0})$,
\begin{eqnarray}
&n(t_{0}) = n_D\Bigg(\dfrac{R_{D}}{R_{0}}\Bigg)^3
\textrm{exp}(-t_{0}/\tau)\nonumber =
\dfrac{\zeta(3)}{\pi^2}~T_{D}^3 \Bigg(\dfrac{R_{D}}{R_{0}}\Bigg)^3
\textrm{exp}(-t_{0} /\tau) \nonumber \\
&= \dfrac{\zeta(3)}{\pi^2}\dfrac{{g_*}_{0}}{{g_*}_{D}}~T_{0}^3
\text{exp}(-t_{0} /\tau) \simeq 7.5 ~\textrm{exp}(-t_{0} /\tau)
~\textrm{cm}^{-3}.
\end{eqnarray}
Note that the coefficient $7.5~\mathrm{cm^{-3}}$ would be
the present dilaton number density if the dilaton had not been
decaying at all, which is half the present number density
of the massless graviton.

\section{Dilaton as a dark matter candidate}

The above analysis implies that the dilaton with a proper mass can
easily survive to present time, and could become the dark matter
of the universe. Assuming this is the case, we can estimate the
mass of the dilaton. It has been argued that there are two mass
ranges of the relic dilaton, $\mu_1 \simeq \textrm{500~eV}$ and
$\mu_2 \simeq \textrm{270~MeV}$, in which the relic dilaton could
be the dominant matter of the universe \cite{cqg98}. This is
because the dilaton with mass larger than $\mu_2$ does not survive
long enough to become the dominant matter of the universe, and the
dilaton with mass smaller than $\mu_1$ survives but fails to be
dominant due to its low mass. The dilaton with mass in between
cannot be seriously considered because it would overclose the
universe. In this section we refine the above result. 

\begin{table}[t]
\begin{tabular}{|c|c|c|c|c|}
\hline
  $g_1\simeq g_2$ & $\mu_1$ & $\tau_1$ & $\mu_2$ & $\tau_2$ \\
  \hline
   10 &~ 160~eV~ &~$3.84 \times10^{33}$~sec &~$75.6 $~MeV  &~$3.62 \times10^{16} $~sec   \\ \hline
    5 &~ 160~eV~ &~$1.53 \times10^{34}$~sec &~$121  $~MeV &~$3.49 \times10^{16} $~sec   \\ \hline
    2 &~ 160~eV~ &~$9.59 \times10^{34}$~sec &~$219 $~MeV  &~$3.35 \times10^{16} $~sec   \\ \hline
    1  &~ 160~eV~ &~$3.84 \times10^{35}$~sec &~$276  $~MeV  &~$3.29 \times10^{16} $~sec  \\
  \hline
   0.9&~ 160~eV~ &~$4.74 \times10^{35}$~sec &~$292 $~MeV  &~$3.27 \times10^{16} $~sec   \\ \hline
  0.8 &~ 160~eV~ &~$6.00 \times10^{35}$~sec &~$312 $~MeV  &~$3.26 \times10^{16} $~sec   \\ \hline
  0.7 &~ 160~eV~ &~$7.83 \times10^{35}$~sec &~$341 $~MeV  &~$3.25 \times10^{16} $~sec   \\ \hline
  0.6 &~ 160~eV~ &~$1.07 \times10^{36}$~sec &~$383 $~MeV  &~$3.21 \times10^{16} $~sec   \\ \hline
  0.5 &~ 160~eV~ &~$1.53 \times10^{36}$~sec &~$445 $~MeV  &~$3.19 \times10^{16} $~sec   \\ \hline
  0.4 &~ 160~eV~ &~$2.40 \times10^{36}$~sec &~$543 $~MeV  &~$3.14 \times10^{16} $~sec   \\ \hline
  0.3 &~ 160~eV~ &~$4.26 \times10^{36}$~sec &~$702 $~MeV  &~$3.09 \times10^{16} $~sec   \\ \hline
  0.2 &~ 160~eV~ &~$9.60 \times10^{36}$~sec &~$988 $~MeV  &~$3.03 \times10^{16} $~sec   \\ \hline
  0.1 &~ 160~eV~ &~$3.84 \times10^{37}$~sec &~$1.68$~GeV  &~$2.94 \times10^{16} $~sec   \\ \hline
  0.05&~ 160~eV~ &~$1.53 \times10^{38}$~sec        &~$2.76 $~GeV  &~$2.85 \times10^{16} $~sec   \\ \hline
  $10^{-2}$ &~ 160~eV~ &~$3.84 \times10^{39}$~sec  &~$8.37 $~GeV  &~$2.66 \times10^{16} $~sec   \\ \hline
  $10^{-3}$ &~ 160~eV~ &~$3.84 \times10^{41}$~sec  &~$40.0 $~GeV  &~$2.45 \times10^{16} $~sec   \\ \hline
  $10^{-4}$ &~ 160~eV~ &~$3.84 \times10^{43}$~sec  &~$191  $~GeV  &~$2.25 \times10^{16} $~sec   \\ \hline

\end{tabular}
\caption{The coupling constants versus dilaton mass
and life-time, where we have assumed $g_1 \simeq g_2$ .
Here the smaller mass is denoted by $\mu_1$
and larger mass is denoted by $\mu_2$, and $\tau_1$ and $\tau_2$
are the life-time of $\mu_1$ and $\mu_2$.}
\end{table}

According to recent cosmological observations,
the dark matter occupies about $23\%$ of the critical density
$\rho_c=3H_{0}^2/(8\pi G) \simeq 10.5~ h^2~ \textrm{keV cm}^{-3}$,
where $h$ is the dimensionless Hubble parameter in units of 100
$\textrm{km~sec}^{-1}~\textrm{Mpc}^{-1}$. On the other hand, the
``dark energy" characterized by the cosmological constant is believed
to occupy about $70\%$ of the total energy of the universe
\cite{teg}. So for the dilaton to be the dark matter of the universe we
must have the following requirement \cite{cqg98},
\bea
&\rho(\mu)=\mu \times 7.5 ~\textrm{exp}[-t_{0} /\tau(\mu)]~
\textrm{~cm}^{-3}=0.23\times \dfrac{3H_{0}^2}{8\pi G} \simeq
0.23\times10.5~ h^2~ \textrm{keV cm}^{-3},
\label{dmass1}
\eea
where $\rho(\mu)$ is the dilaton mass density. At the same time, 
the energy density $\tilde \rho (\mu)$ of the daughter particles 
(photons and light fermions) coming from the dilaton decay should 
be negligible compared to the critical density. This gives
the second requirement
\bea
&\tilde \rho (\mu) \ll \rho_c.
\label{dmass2}
\eea
To find the dilaton mass which satisfies these constraints,
we have to know the coupling constants
$g_1$ and $g_2$.  In Kaluza-Klein unification they are
given by \cite{prd87}
\bea
g_1=\sqrt{\dfrac{n+2}{n}},~~~~~g_2=\sqrt{\dfrac{n}{n+2}}.
\eea
But in the following we will leave them as free parameters,
although our favorite values are $g_1 \simeq g_2 \simeq 1$.
Now, with $t_0=1.5\times10^{10}~\textrm{yr} =4.73\times
10^{17}~\textrm{sec}$ and $h\simeq 0.7$, we obtain the numerical
solutions of the first constraint (\ref{dmass1}) shown in TABLE I. 
As we see in the table, it has two solutions
for the dilaton mass and life-time for given coupling
constants. We denote the smaller one by $\mu_1$ and $\tau_1$ and
the larger one by $\mu_2$ and $\tau_2$ in the table.  
In our numerical calculations, the decay
channels we considered are $\gamma\gamma$, $\nu\overline{\nu}$,
$e^+e^-$, $\mu^+\mu^-$ processes. So when $g_1 \simeq
g_2 \gtrsim 5\times 10^{-2}$, our calculations are exact. But
when $g_1 \simeq g_2 \lesssim 10^{-2}$, the dilaton
has larger mass and can
decay into other heavier particles like $\tau^+ \tau^-$.
But even in the latter case, the two-photon decay probability
is far greater than the fermion-antifermion decay probability
except when $m \simeq 0.32\times \mu$ (in which case we have
$\Gamma_{\sigma\rightarrow\bar{\psi}\psi} \simeq 1.49 \times
\Gamma_{\sigma\rightarrow\gamma\gamma}$) as we have remarked,
and the error in evaluating the dilaton mass
in the latter case is at most $20\%$ or so.

\begin{figure}
\epsfig{file=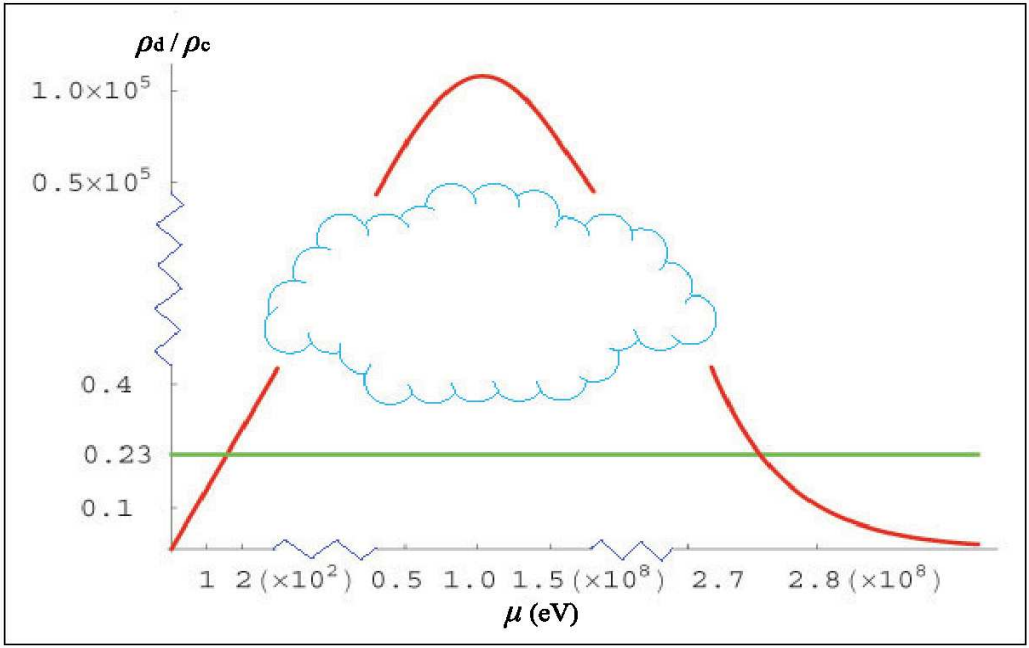, height = 9cm, width = 14cm}
\caption{\label{Fig. 1} The dilaton mass density $\rho(\mu)$
versus the dilaton mass $\mu$,
obtained with $g_1 \simeq g_2 \simeq 1$.}
\end{figure}

Note that the smaller mass $\mu_1$ is insensitive to
the values of the coupling constants, while the larger mass
$\mu_2$ increases as the coupling constants decrease. On the other
hand, the life-time $\tau_1$ is sensitive to the values of the
coupling constants, while the life-time $\tau_2$ remains of the
same order for all values of the coupling constants. 

With $g_1 \simeq g_2 \simeq 1$ we can plot the dilaton density
$\rho(\mu)$ against its mass $\mu$, which is shown in Fig.1.
Note that $\rho(\mu)$ starts from zero and approaches to 
the maximum value of about $1.08 \times 10^{5} \rho_c$ at 
$\mu \simeq 103~{\rm MeV}$, and
again decreases to zero when $\mu$ goes to infinity.
More importantly, $\rho(\mu)$ exceeds the dark matter density 
in the range $160 ~\textrm{eV} < \mu < 276 ~\textrm{MeV}$. 
This means that when $ \mu < 160 ~\textrm{eV}$ or 
$\mu > 276 ~\textrm{MeV}$, the dilaton undercloses the universe, 
but when $160 ~\textrm{eV} < \mu < 276 ~\textrm{MeV}$ 
it overcloses the universe.
This immediately rules out the dilaton with mass range
$160 ~\textrm{eV} < \mu < 276 ~\textrm{MeV}$.
Moreover, we have two possible mass ranges which are of 
particular interest, $\mu_1\simeq 160 ~\textrm{eV}$ with 
life-time $\tau_1 \simeq 3.84 \times10^{35}~\textrm{sec}$ and 
$\mu_2\simeq 276 ~\textrm{MeV}$ with life-time $\tau_2 \simeq
3.29\times10^{16}~\textrm{sec}$, which makes the dilaton 
the dominant matter of the universe.

\begin{table}[t]
\begin{tabular}{|c|c|c|c|c|}
\hline
  $\rho_d/\rho_c$ & $\mu_1$ & $\tau_1$ & $\mu_2$ & $\tau_2$ \\  \hline
  $100~\%$ &~ 686~eV~ &~$4.85 \times10^{33}$~sec &~$270 $~MeV  &~$3.65 \times10^{16} $~sec   \\ \hline
   $23~\%$ &~ 160~eV~ &~$3.84 \times10^{35}$~sec &~$276 $~MeV  &~$3.29 \times10^{16} $~sec   \\ \hline
   $10~\%$  &~ 68.6~eV~ &~$4.85 \times10^{36}$~sec &~$280  $~MeV &~$3.12 \times10^{16} $~sec   \\ \hline
   $4~\%$  &~ 27.4~eV~ &~$7.58 \times10^{37}$~sec &~$284 $~MeV  &~$2.93 \times10^{16} $~sec   \\ \hline
   $1~\%$  &~ 6.86~eV~ &~$4.85 \times10^{39}$~sec &~$291  $~MeV  &~$2.68 \times10^{16} $~sec  \\  \hline
   $0.5~\%$  &~3.43~eV~ &~$3.88 \times10^{40}$~sec &~$294  $~MeV  &~$2.59 \times10^{16} $~sec  \\  \hline
\end{tabular}
\caption{The dilaton mass and life-time versus the ratio
$\rho_d/\rho_c$. Here the coupling constants $g_1$ and $g_2$
are set to be 1.}
\end{table}

So far we have assumed that the dilaton occupies all of the
dark matter, about 23\% of the critical density $\rho_c$. But
even when we loosen this constraint, we get 
similar result. Varying the ratio $\rho_d/\rho_c$ of the
dilaton's mass density to the critical density,
we obtain the result shown in TABLE II with
$g_1\simeq g_2 \simeq 1$. The result shows that $\mu_1$ and $\tau_1$
are sensitive to the change of $\rho_d/\rho_c$, but $\mu_2$ and $\tau_2$
are not much affected by that. Moreover,
the generic feature of the dilaton physics remains the same.

Now, we have to make sure that the dilaton mass should 
also satisfy the second constraint (\ref{dmass2}). To check this,
notice that the $160~{\rm eV}$ dilaton is almost stable
because $\tau_1 \simeq 8.1 \times 10^{17}~t_0$. 
So the energy density of the daughter particles 
must be negligible compared to the energy density of the dilaton.
This means that this dilaton can easily satisfy 
the second constraint (\ref{dmass2}). 
On the other hand, most of the $276~{\rm MeV}$ dilaton 
should have decayed by now, because
$\tau_2 \simeq 6.9 \times 10^{-2}~t_0$. Indeed only 
$0.5 \times 10^{-6}$ of the heavy dilaton survives now. 
So the energy density of the daughter particles 
becomes much bigger than that of the dilaton. This means that 
the daughter particles from the heavy dilaton overclose
the universe, and thus can not satisfy the second constraint. 
This effectively rules out the heavy dilaton. 
So only the $160~{\rm eV}$ dilaton can be accepted as 
the dark matter candidate.

\begin{figure}
\epsfig{file=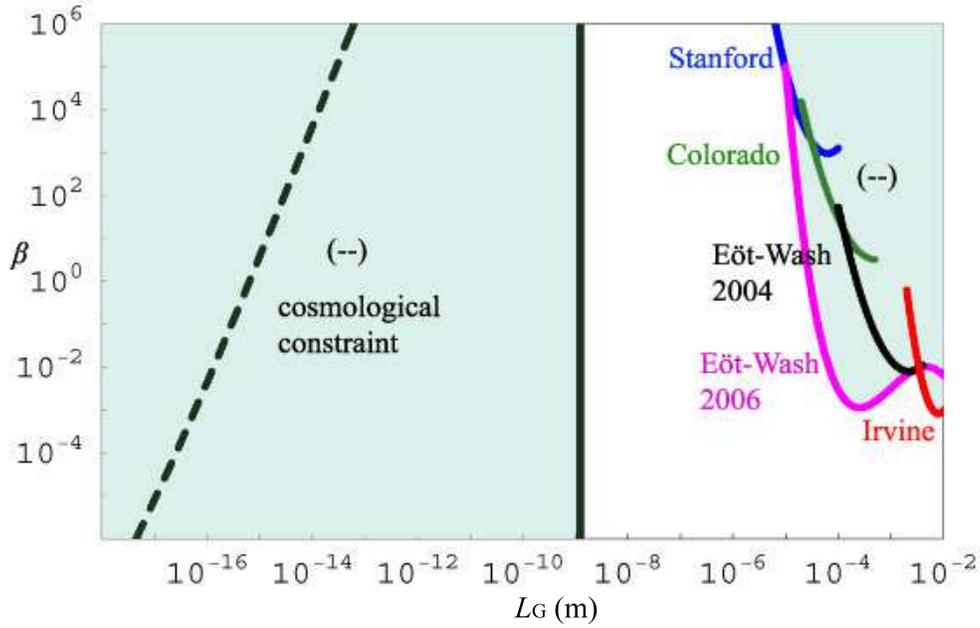, height = 9cm, width = 14cm}
\caption{\label{Fig. 2} The possible scale $L_G=1/\mu$
of the internal space versus the relative fine structure
constant $\beta=\alpha_5/\alpha_g$ of the fifth force
in $(4+3)$-dimensional unification
with $S^3$ compactification of the internal space.
The colored region marked by (--)
is the excluded region, and the dotted line represents the constraint
of the heavy dilaton whose daughter particles overclose the universe.}
\end{figure}

The dark matter dilaton has the following characteristics. 
With the mass $\mu\simeq 160 ~\textrm{eV}$, the possible decay
channels of the dilaton are the $\gamma\gamma$ and three
$\nu\overline{\nu}$ processes. But with life-time  
$\tau \simeq 8.1 \times 10^{17}~t_0$ this dilaton is almost stable.
To see whether this can be hot or cold dark matter,
we should estimate the free-streaming distance $\lambda_{FS}$
of the dilaton first. The
dilaton in this case becomes nonrelativistic at $T_{NR} \simeq
\mu/3 \simeq 53.3 ~\textrm{eV} $ well before the
matter-radiation equilibrium era $t_{EQ}\simeq 4.36\times10^{10}
(\Omega_0 h^2)^{-2}~\textrm{sec}\simeq 1.82\times
10^{11}~\textrm{sec}$. The time $t_{NR}$ when the
dilaton becomes nonrelativistic is given by \cite{kol}
\bea
&t_{NR}\simeq 1.2\times10^7 \Bigg( \dfrac{\textrm{keV}}{\mu}
\Bigg)^2 \Bigg( \dfrac{T_{NR}}{T_{\gamma}} \Bigg)^2
\textrm{sec}=1.2\times10^7 \Bigg( \dfrac{\textrm{keV}}{\mu}\Bigg)^2
\Bigg(\dfrac{g_{*NR}}{g_{*D}} \Bigg)^{2/3} \textrm{sec}, \nn\\
&\dfrac{t_{EQ}}{t_{NR}}=\Big[\dfrac{\mu/\textrm{eV}}{17(\Omega_0
h^2)(T_{NR}/T_{\gamma}) } \Big]^2,
\eea
where $T_{\gamma}$ is the temperature of the photon at $t_{NR}$,
$g_{*NR}$ is the total relativistic degrees of freedom when the
dilaton becomes non-relativistic. So the free-streaming distance
$\lambda_{FS}$ is given by \cite{cqg98,kol},
\bea
&\lambda_{FS}\simeq 0.2\textrm{Mpc}\bigg(
\dfrac{\mu}{\textrm{keV}} \bigg)^{-1} \bigg(
\dfrac{T_{NR}}{T_\gamma} \bigg)~ \Big[
\textrm{ln}\bigg(\dfrac{t_{EQ}}{t_{NR}} \bigg)+2 \Big] \nonumber \\
&=0.2\textrm{Mpc}\bigg( \dfrac{\mu}{\textrm{keV}} \bigg)^{-1}
\bigg(\dfrac{g_{*NR}}{g_{*D}}  \bigg)^{1/3}~ \Big[
\textrm{ln}\bigg(\dfrac{t_{EQ}}{t_{NR}} \bigg)+2 \Big].
\eea
Now, with $g_{*D}\simeq~106.75$ and $g_{*NR}\simeq 3.91$
we get $t_{NR}\simeq 5.17\times10^7 ~\textrm{sec}$ and
$\lambda_{FS}\simeq 4.2 ~\textrm{Mpc}$. Comparing the latter
with the typical structure formation scale $\lambda_{EQ} \simeq
13(\Omega_0 h^2)^{-1} \simeq 18.6 ~\textrm{Mpc}$, we
may conclude that the 160~\textrm{keV} dilaton becomes
a warm dark matter. 

In comparison the dilaton with mass $\mu\simeq 276 ~\textrm{MeV}$
becomes non-relativistic at $T_{NR} \simeq \mu/3 \simeq 92
~\textrm{MeV} $. The decay channels available here are
$\gamma\gamma$, $\nu\overline{\nu}$, $e^+e^-$, $\mu^+\mu^-$
processes. Among them, only the $\mu^+\mu^-$ process is comparable
to the $\gamma\gamma$ process since the mass of the muon
$m_{\mu}\simeq 106~\textrm{MeV}$ is around $0.32\times\mu_2$. 
With life-time $\tau \simeq 6.9 \times 10^{-2}~t_0$, 
only a fraction of the dilatons have survived up to now. In this case
we have $g_{*NR}\simeq 19.5$ since only photon, three neutrinos,
electron, and muon could be in thermal equilibrium at $t=t_{NR}$.
With this value, we get $t_{NR}\simeq 5.07\times10^{-5}
~\textrm{sec}$ and $\lambda_{FS}\simeq 1.55\times 10^{-5}
~\textrm{Mpc} \ll \lambda_{EQ} \simeq 18.6 ~\textrm{Mpc} $. So
this dilaton could have been an excellent 
candidate for cold dark matter. But of course, this dilaton 
is not acceptable because the daughter particles 
overclose the universe.

As we have shown there are two constraints on the dilaton mass,
the experimental constraint from the fifth force and the
theoretical constraint from cosmology. Clearly these constraints
restricts the allowed scale of the internal space. Putting the two
constraints together we obtain Fig.2, which shows the allowed
regions of the scale of the internal space versus 
the relative fine structure constant 
$\beta=\alpha_5/\alpha_g$ of the fifth force. 
Notice that the cosmological constraint tells that 
the scale of the internal space can not be smaller than
$10^{-9}~{\rm m}$.

\section{Dilaton Detection Experiment}

So far, we have tried to estimate the dilaton mass based on the
conjecture that the dilaton is the dark matter of the universe.
Now an important question is how to detect the relic dilaton and
confirm such conjecture. Clearly one could try to establish the
existence of the dilaton measuring the dilatonic fifth force
\cite{grg91,exp1}. But the above analysis implies that,
if indeed the dilaton is the dark matter of the universe,
it's detection by the fifth force experiments would be
almost impossible because such dilaton
generates an extremely short ranged fifth force.

In this section we propose a totally different
type of experiment based on two photon decay of the relic dilaton.
Of course, one might try to detect the two photon decay of the
relic dilaton directly, searching for the mono-energetic x-ray
signals from the sky \cite{cqg98}.
Here we propose another type of experiment, a Sikivie-type experiment
which detects the dilaton conversion to one photon in strong
electromagnetic background. In this type of experiment
the dilaton conversion rate can be greatly enhanced by two factors,
first by the strong electromagnetic background and
secondly by the large dilaton density of halo.
It is clear that the conversion rate is enhanced by
the strong background, because the conversion amplitude 
is proportional to the background field strength. 
Moreover, just as in the axion detection experiment,
we can assume that our galaxy halo
is made of the relic dilaton if the dark matter is the dilaton.
In this case the conversion rate will be enhanced by a
factor $10^{5}$, because the average energy density of the relic
dilaton $0.23\times10.5~ h^2~ \textrm{keV cm}^{-3}\simeq
1.18~\textrm{keV cm}^{-3}$ in the present universe can be replaced
by the galaxy halo density $\rho_{halo} \simeq 0.3~ \mathrm{GeV
cm^{-3}}$ \cite{asz}. In the following we estimate the power of dilaton
conversion to one photon in strong magnetic background,
assuming that our galaxy halo is made of dilaton.

Consider a rectangular cavity with
three edges $L_x,~L_y,~L_z$ and volume $V=L_x L_y L_z$
made of a perfect conductor, which has a
strong magnetic background
$\vec{B}_{ext}(\vec{x})=B_{ext}(\vec{x})~\hat{z}$ in the
z-direction inside, and consider the halo dilaton conversion in
the cavity described by the interaction
\begin{eqnarray}
\mathcal{L}_{\sigma\gamma\gamma}=-\alpha_1 \hat{\sigma}
F_{\mu\nu}^2=2\alpha_1\hat{\sigma}(\vec E^2-\vec B^2),
~~~~~~\alpha_1=\dfrac{1}{4} g_{1}\sqrt{16\pi G}.
\end{eqnarray}
In this case the induced photon is
described by TE mode (the magnetic wave)
$\vec{B}(\vec{x})=B(\vec{x})~\hat{z}$, and the differential
cross-section $d\sigma$ of the dilaton conversion in the cavity is
given by
\bea
&d\sigma_{\vec{k},\lambda}=2\pi\delta(k^{0}-p^{0})
\dfrac{1}{2p^{0}v} \dfrac{d^{3}\vec{k}}{(2\pi)^{3}2k^{0}}|M|^{2}, \nn\\
&M =-i 4 \alpha_1 \vec{B}(\vec{x})\cdot \vec{B}_{ext}(\vec{q})
=-i 4 \alpha_1 k^{0}(\hat{\epsilon}(\vec{k},\lambda)\times
\hat{k}) \cdot \vec{B}_{ext}(\vec{q}),
\eea
where $p^\mu$ and $k^\mu$ are the 4-momenta of the dilaton
and the induced photon, $M$ is the Feynman reduced matrix element,
$\hat{\epsilon}(\vec{k},\lambda=\pm1)$ and $\hat{k}$ are
the 3-dimensional photon polarization vector and the unit vector
in the direction of the photon momentum $\vec{k}$,
$\vec{q}=\vec{k}-\vec{p}$ is the spatial momentum transfer,
and $\vec{B}_{ext}(\vec{q})$ is the Fourier transform of
$\vec{B}_{ext}(\vec{x})$. Note that in the classical background
only energy is conserved, and the $\delta(k^{0}-p^{0})$ term
represents this fact. Then the total cross-section $\sigma$ in the
continuum limit is given as follows,
\bea
&\sigma=\dfrac{}{} \sum_{\lambda=\pm 1} 2\pi \int d^{3}\vec{k}
~\delta(k^{0}-p^{0}) \dfrac{1}{2p^{0}v}
\dfrac{1}{(2\pi)^{3}2k^{0}}|M|^{2}\nonumber \\
&=\dfrac{\alpha_1^{2}}{\pi^{2} v }\sum _{\lambda=\pm 1} \int
d^{3}\vec{k} \delta(k^{0}-p^{0}) |\hat{B}(\vec{k},\lambda) \cdot
\vec{B}_{ext}(\vec{q})|^{2},
\eea
where $\hat{B}(k,\lambda)=\hat{\epsilon}(\vec{k},\lambda)\times
\hat{k}$ is the unit vector in the direction of
the induced magnetic field $\vec{B}$.

Let the wave vector of the photon be
$\vec k=(n_x \pi/L_x,~n_y\pi/L_y,~n_z \pi/L_z)$ where
$(n_x,~n_y,~n_z)$ are arbitrary
integers. For TE modes, the boundary condition
\begin{eqnarray}
B~(z=0,L_{z})=0~,~~\dfrac{\partial B}{\partial x}(x=0,
L_{x})=0~,~~ \dfrac{\partial B}{\partial y}(y=0, L_{y})=0,
\end{eqnarray}
requires the induced magnetic field to assume the form
\begin{eqnarray}
B=A\cos\bigg(\dfrac{n_{x}\pi x}{L_{x}}\bigg)
\cos\bigg(\dfrac{n_{y}\pi y}{L_{y}}\bigg)
\sin\bigg(\dfrac{n_{z}\pi z}{L_{z}}\bigg),
\end{eqnarray}
where $A$ is a normalization constant.
Notice that $n_{x}$ and $n_{y}$ cannot be zero simultaneously,
and $n_{z}$ must be a non-zero integer \cite{jac}.

Now, we have
\bea
&\dfrac{}{}\sum _{\lambda=\pm 1}|\hat{B}(\vec{k},\lambda) \cdot
\vec{B}_{ext}(\vec{q})|^{2}
=|\hat{k} \times B_{ext}(\vec{q})\hat{z}|^{2}
=\dfrac{k_{x}^2+k_{y}^2}{{(k^0)}^2}|B_{ext}(\vec{q})|^{2}, \nn\\
&\vec{B}_{ext}(\vec{q})=\dfrac{}{}\int_{V} e^{i\vec{q}\cdot
\vec{x}}\vec{B}_{ext}(\vec{x})~d^3 \vec{x}
=\dfrac{}{} \int_{V} e^{i\vec{q}\cdot \vec{x}}B_{ext}(\vec{x})\hat{z}~ d^3
\vec{x}= B_{ext}(\vec{q})\hat{z},
\eea
so that, changing the integration into summation as follows
\begin{eqnarray}
d^{3}\vec{k} &=& dk_{x}dk_{y}dk_{z}= \frac{\pi}{L_{x}}
\frac{\pi}{L_{y}}
\frac{\pi}{L_{z}}dn_{x}dn_{y}dn_{z}=\frac{\pi^{3}}{V},
\end{eqnarray}
we get the following cross-section
\begin{eqnarray}
\sigma&=& \sum _{\vec{k}}\frac{\pi \alpha^{2}}{Vv}
\frac{(k_{x}^{2}+k_{y}^{2})}{(k^0)^2}
\delta(k^0-p^0)|B_{ext}(\vec{q})|^{2}.
\end{eqnarray}
To proceed, we let
\bea
\vec{B}_{ext}(\vec{x})=B_{0} \cos(Qx)\hat{z},
\label{rext}
\eea
and approximate $\vec{q}=(\vec{k}-\vec{p}) \sim
\vec{k}$ since the incoming halo dilaton is highly non-relativistic
(with $v \sim 10^{-3}c) $ \cite{asz}. In this case we have
\bea
&|B_{ext}(\vec{q})|^{2}=|\dfrac{}{}\int_{V}d^{3}\vec{x}e^{i\vec{q}
\cdot \vec{x}} B_{0} \cos(Qx)|^{2} \nn\\
&=B_{0}^{2}L_{x}^{2}L_{y}^{2}L_{z}^{2}
\dfrac{\sin^{2}(\dfrac{k_{y}L_{y}}{2})}
{(\dfrac{k_{y}L_{y}}{2})^{2}}
\dfrac{\sin^{2}(\dfrac{k_{z}L_{z}}{2})}
{(\dfrac{k_{z}L_{z}}{2})^{2}} \times \dfrac{1}{4}\Big(\Big[
\dfrac{\sin(k_{x}-Q)L_{x}}{(k_{x}-Q)L_{x}}
+\dfrac{\sin(k_{x}+Q)L_{x}}{(k_{x}+Q)L_{x}} \Big]^{2} \nn\\
&+\Big[ \dfrac{1-\cos(k_{x}-Q)L_{x}}{(k_{x}-Q)L_{x}}
+\dfrac{1-\cos(k_{x}+Q)L_{x}}{(k_{x}+Q)L_{x}} \Big]^{2} \Big).
\eea
As we can see, $|B_{ext}(\vec{q})|^{2}$ has the maximum value
\begin{eqnarray}
|B_{ext}(\vec{q})|^{2}_{max}=\frac{B_0^{2}L_x^{2}L_y^{2}L_z^{2}}{\pi^2}~,
\end{eqnarray}
when
\begin{eqnarray}
k_x=\pm Q, ~k_zL_z=\pm \pi~, ~k_yL_y=0~.
\end{eqnarray}
Note that $|B_{ext}(\vec{q})|^{2}_{max}$ would be
highly suppressed without the external sinusoidal background,
which is why we choose the sinusoidal external
magnetic field (\ref{rext}).

We are interested in the dilaton with the mass range $\mu \gtrsim
0.1 ~\textrm{keV}$ whose Compton wave-length is of order smaller
than $2\times 10^{-7}~\textrm{cm}$. Considering the typical detector
length scale $L_x,~L_y,~L_z \simeq 1~\mathrm{m}$ and
$(k^0)^2=k_{x}^{2}+k_{y}^{2}+k_{z}^{2}\simeq \mu^2$, we have $
k_{x} \simeq k^0 \gg {\rm Max}~(k_y, k_z)$ since $\mu L_x,~\mu L_y \gg 1$ in
the resonance case. Thus we can use the following approximation
\begin{eqnarray}
k^0dk^0=k_xdk_x+k_ydk_y+k_zdk_z \simeq k_xdk_x \Rightarrow
dk^0=\frac{k_x}{k^0}dk_x\simeq dk_x~.
\end{eqnarray}
On the other hand, the number of additional modes due to the
differential spread $dk^0$ around $k^0=p^0$ is
\begin{eqnarray}
&dn_x = \dfrac{L_x}{\pi}dk_x  \simeq \dfrac{L_x}{\pi} dk^0,
~~~~~\delta(k^0-p^0) dn_x=\dfrac{L_x}{\pi}~.
\end{eqnarray}
Combining these relations, we finally obtain
\bea
&\sigma= \dfrac{}{}\sum _{\vec{k}}\dfrac{\pi \alpha_1^{2}}{Vv}
\dfrac{(k_{x}^{2}+k_{y}^{2})}{(k^0)^2}
\delta(k^0-p^0)|B_{ext}(\vec{q})|^{2}
\simeq \dfrac{4\pi \alpha_1^{2}}{Vv} \dfrac{(k^0)^2}{(k^0)^2}
\delta(k^0-p^0)dn_x \times
|B_{ext}(\vec{q})|^{2}_{max} \nonumber \\
&= \dfrac{16\pi \alpha_1^{2}}{Vv}\dfrac{L_x}{\pi} \times
|B_{ext}(\vec{q})|^{2}_{max}
= \dfrac{4\alpha_1^{2}}{\pi^2Vv}B_0^{2}L_x^{2}L_y^{2}L_z^{2}L_x
=\dfrac{4\alpha_1^{2}}{\pi^2v} B_0^{2}V L_x,
\eea
and the following detection power $P$
\begin{eqnarray}
&P=\mu~n_{d}v\sigma=\Bigg(\dfrac{4\alpha_1^{2}}{{\pi}^{2}}\Bigg)
\rho_{d}B_{0}^{2}L_{x}V,
\end{eqnarray}
where $n_{d}$ is the dilaton number density and $\rho_{d}$ is the
dilaton energy density. Notice that the detection power
depends on the energy density, not the mass, of dilaton.

This agrees with that of the axion detection power except for the
numerical factor of order unity which comes from the different
axion-photon coupling constant. In the case of the axion, the
axion-photon interaction Lagrangian and axion detection power 
are given as follows \cite{sik},
\begin{eqnarray}
&\mathcal{L}_{a\gamma\gamma}=-\alpha_{\gamma}~a
~F_{\mu\nu}\tilde{F}^{\mu\nu}=4\alpha_{\gamma}~a~\vec{E}\cdot\vec{B},\nn\\
&P_a= 2\alpha_{\gamma}^2 \rho_{a}B_{0}^{2}L_{x}V.
\end{eqnarray}
As we have mentioned there are two types of axion, the popular
axion from strong interaction and the gravitational axion proposed
as a pseudoscalar graviton \cite{pec,ni}. The difference is that
for the popular axion the coupling constant $\alpha_{\gamma}$ is
given by $g_\gamma\alpha/4\pi f_a$, where $g_\gamma$ is a
model-dependent dimensionless coupling constant of order one,
$\alpha$ is the electromagnetic fine-structure constant, and $f_a$
is the $U_{PQ}(1)$ symmetry breaking scale. But for the
gravitational axion $\alpha_{\gamma}$ is similar to our $\alpha_1$
because this axion is the pseudoscalar partner of the dilaton.
Other than this they are virtually identical.

We can compare the axion detection power with the dilaton
detection power. Consider the popular axion first. Since $f_a$ is
related to the axion mass $m_a$ by $m_a\simeq 6~\mathrm{eV}\times
10^6~\mathrm{GeV}/f_a$, and the educated guess of the axion mass
is around $10^{-6}~\mathrm{eV}$ or so, we have $ f_a \simeq 6
\times 10^{12}~{\rm GeV}$ \cite{sik,asz}. So we have
\bea
&\dfrac{P}{P_a}\simeq 1.9\times 10^{6} \times
\big(\dfrac{g_1}{g_\gamma} \big)^2\big(\dfrac{f_a}{m_p} \big)^2
\simeq 4.7 \times 10^{-7}.
\eea
This is a small number, but this is because $\alpha_1$
is smaller than $\alpha_\gamma$ due to the fact that
$f_a$ is much bigger than the Planck mass.
Indeed, with $\rho_{d}\simeq\rho_{halo}\simeq0.5 \times 10^{-24}~
\mathrm{g/cm^{3}}\simeq 0.3 ~\mathrm{GeV/cm^{3}}$ and $B_{0}=
10~\mathrm{Tesla}$, $L_{x}=L_{y}=L_{z}\simeq 1~\mathrm{m}$, we get
the dilaton detection power $P\simeq 1.42 \times 10^{-31}~\textrm{W}$
with $g_1 \simeq 1$. This is
$10^{-5}$ times smaller than the axion detection power
in current experiments \cite{sik,asz}.
For the gravitation axion, however, we expect 
$\alpha_1 \simeq \alpha_\gamma$ so that $P_a$ becomes 
as small as $P$.
So in this case the axion detection power becomes smaller than
the popular axion detection power, and becomes comparable
to the dilaton detection power.

Notice that, due to the pseudoscalar coupling,
the axion produces TM modes (the electric wave)
rather than TE modes. Another notable difference
between the dilaton and the axion is that
for the dilaton the photon polarization is perpendicular to the
external magnetic field, whereas for the axion
the photon polarization is parallel to the
external magnetic field.

Now, a few remarks are in order. First, the above result holds
when we have the resonance, $Q \simeq \mu$. But it seems very
difficult to make static magnetic field of wavelength of order
$\mu ^{-1} \lesssim 10^{-7}~\mathrm{cm}$ with the current
technology. However we may be able to set up x-ray range
electromagnetic waves with $\omega_{ext} = Q \simeq \mu$. In that
case, the only change needed is to replace $\delta(k^{0}-p^{0})$
by $\delta(k^{0}-p^{0}\pm p^{0})$ in the above calculation, which
will make the detection power $P$ twice as big. Second, the
dilaton detection power appears too small to be considered
realistic at present. On the other hand, we notice that the
relevant technologies are developing fast \cite{asz}, so that it
may be possible to detect the halo dilaton in the near future.
Third, we have used the magnetic background in the above
calculation. With an electric background the detection power would
have been proportional to the electric field energy density. In
terms of the field energy density, 1 Tesla corresponds to
$300~\mathrm{MV/m}$ since $E=cB$ in the MKS unit system. But the
strongest magnetic field and electric fields currently available
are around 50 Tesla and 40 MV/m \cite{MEmax}, respectively. So at
present a magnetic background can give us larger detection power.
Moreover, in the air the electric breakdown happens when the
electric field is about 3 MV/m. This is why we have used the
magnetic background in our calculation. And this is why one hardly
uses an electric background in particle creation or annihilation
experiments in laboratories.

\section{Discussion}

The Newton's constant in Einstein's theory has always been a
mystery. The Einstein's theory has a dimensional coupling 
constant, the Newton's constant, because the source of gravity 
is the energy-momentum tensor. The problem is that 
in mass scale, this coupling constant is absurdly bigger than the
ordinary elementary particle mass scale. In this paper we have
shown how the dilaton from higher-dimensional unification can
naturally resolve this mystery. First, the dilaton makes the
Newton's constant a space-time dependent parameter. This changes
the hierarchy problem from a fundamental problem to a space-time
dependent artifact. Moreover, it reduces the Planck mass down to
the ordinary mass scale when the internal space becomes larger
than the Planck size. This is because the dilaton mass is fixed by
the curvature of the internal space. So it can be small even
though the unit of the curvature is set by the Planck mass. When the
internal space becomes larger, it becomes flatter and the
curvature becomes smaller. This means that the dilaton mass can be
much smaller than the Planck mass when the size of the internal
space is big enough. This is how the Kaluza-Klein dilaton resolves
the hierarchy problem \cite{cho}.

As the scalar graviton the dilaton couples to all matters, so that
it creates the fifth force which modifies Einstein's gravity. This
is why the fifth force experiments have been used to detect the
dilaton. On the other hand the dilaton coupling to matter fields
is very weak. This means that the dilaton can easily survive to
present universe. This makes the dilaton an excellent candidate of
dark matter. Our analysis tells that there is practically 
only one mass range, $\mu \simeq 160~{\rm eV}$, for which 
the dilaton can be the dark matter. This cosmological
constraint of dilaton mass implies that detecting the dilaton by
the fifth force experiments would be futile, because the fifth
force is too short ranged to be detected in the near future.

In this paper we have proposed a totally different type of
experiment to detect the dilaton, based on the dilaton photon
conversion in strong magnetic background. Although the detection
power of dilaton is still very small, we hope that this type of
experiment could help us to detect the dilaton.

{\bf ACKNOWLEDGEMENT}

~~~The work is supported in part by the BSR Program (Grant
KRF-2007-314-C00055) of Korea Research Foundation and by the ABRL
Program (Grant R14-2003-012-01002-0) of
Korea Science and Engineering Foundation.

\end{document}